\documentclass[a4paper,fleqn,usenatbib,useAMS]{mnras}

\usepackage[dvips]{graphicx}
\usepackage{color}
\usepackage{times}
\usepackage{natbib}
\usepackage{setspace}
\usepackage{amsmath}
\usepackage{amsfonts}
\usepackage{amssymb}
\usepackage{multirow}
\usepackage{multicol}
\usepackage{aas}
\usepackage{booktabs}
\usepackage{adjustbox}
\usepackage{float}
\usepackage{graphics}

\usepackage[applemac]{inputenc}

\def\beq{\begin{equation}}
\def\eeq{\end{equation}}

\def\modot{M$_\odot$}

\def\psion{$\xi_{ion}$}

\def\Qint{Q$_{\rm ion}^*$}
\def\Lya{Ly$\alpha$}
\def\EWlya{EW(Ly{$\alpha$})}
\def\fesclyc{$f{_{\rm esc,LyC}}$}  
\def\fesclya{$f{_{\rm esc,Ly\alpha}}$} 
\def\Herg{erg$^{-1}$ Hz}

\title[Reionisation of the BDF at z\,$\sim 7$] {Is the Bremer Deep Field reionised, at z$\sim$ 7?}

\author[J.M.~Rodr\'i{}guez Espinosa et al.]{J.M. Rodr\'i{}guez Espinosa$^{1,2},$\thanks{E-mail: jre@iac.es} J.M. Mas-Hesse$^3$, R. Calvi$^{1,2}$
\\
$^1$Instituto de Astrof\'isica de Canarias, E-38205 La Laguna, Spain \\ 
$^2$Depto. de Astrof\'isica, Universidad de La Laguna, E-38206 La Laguna, Spain\\
$^3$Centro de Astrobiolog\'ia (CSIC-INTA), Depto. de Astrof\'isica, Madrid, Spain \\
}

\pubyear{2021}

\begin{document}

\label{firstpage}
\pagerange{\pageref{firstpage}--\pageref{lastpage}}
\maketitle

\begin{abstract}
  We show herein that the population of star forming galaxies in the Bremer Deep Field (BDF) have enough ionising power to form two large ionised bubbles which could be in the process of merging into a large one with a volume of 14000 cMpc$^3$. The sources identified in the BDF have been completed with a set of expected low luminosity sources at z $\sim$ 7. We have estimated the number of ionising photons per second produced by the different star forming galaxies in the BDF. This number has been compared with the number that would be required to ionise the bubbles around the two overdense regions. We have used, as reference, ionising emissivities derived from the AMIGA cosmological evolutionary model. We find that even using the most conservative estimates, with a Lyman continuum escape fraction of 10$\%$ the two regions we have defined within the BDF would be reionised. Assuming more realistic estimates of the ionising photon production efficiency, both bubbles would be in the process of merging into a large reionised bubble, such as those that through percolation completed the reionisation of the universe by z = 6. The rather small values of the escape fraction required to reionise the BDF are compatible with the low fraction of faint Ly$\alpha$ emitters identified in the BDF. Finally, we confirm that the low luminosity sources represent indeed the main contributors to the BDF ionising photon production.
\end{abstract}

\begin{keywords}
cosmology: dark ages, reionisation, first stars;  galaxies: starburst; galaxies: high-redshift; 
\end{keywords}

\section{Introduction}
\label{intro}
Many high redshift sources are known from various surveys made in the past decades. Most of the detections have been done through broad band searches \citep{Stark2010, Bouwens2010, Bouwens2006, Steidel2005}, but also with narrow band filters tuned to the Lyman\,$\alpha$ line \citep{Kritt2017, Ouchi2008, Ouchi2010}. However the number of spectroscopic confirmations of the sources detected in the various surveys is rather scarce \citep{Calvi2019, Castellano2018, Harikane2018}. In particular, \citet{Vanzella2011} discovered spectroscopically two sources in the Bremer Deep Field (BDF) at redshift 7. Furthermore, \citet{Castellano2018} detected an additional source at a similar redshift within the same field.  

There are quite a few high-z proto-clusters discovered in the past few years \citep{Castellano2016,  Chanchaiworawit2018, Toshikawa2012, Harikane2018,Oteo2018,Abdullah2018, Jiang2018,Higuchi2019, Harikane2019}. Their importance in the reionisation process has indeed been recognised. Interestingly, the relative fraction of volume occupied by proto-clusters increases with $z$ \citep{Chiang2017}. Besides, proto-clusters represent a collection of sources that together could produce sufficient ionising photons, such that a fraction of them could escape contributing to the reionisation of the universe through the creation of ionised bubbles.  

Papers discussing the presence of ionised bubbles are appearing more often lately, such as \citet{Tilvi2020} that shows evidence of a bubble ionised by 3 Lyman alpha emitting galaxies at $z = 7.7$. Also, \citet{Meyer2020} have found a double peaked Ly$\alpha$ source producing its own ionised bubble at $z = 6.8$. We have also recently characterised an ionised bubble powered by a proto-cluster at $z \sim 6.5$ \citep{RodriguezEspinosa2020}. Finally, \citet{Castellano2018} have explored the conditions around the three Ly$\alpha$ Emitters (LAE) they had identified in the Bremer Deep Field (BDF). While the detection of three LAE's at $z \sim 7$ would point to the presence of a large enough re-ionised bubble allowing for Ly$\alpha$ to escape, the non-detection of any Ly$\alpha$ emission from other 14 Lyman Break Galaxies (LBG) identified in the field was apparently at odds with this scenario.  \citet{Castellano2018}  analysed the possibility that the three Ly$\alpha$ Emitters (LAE) in the BDF could produce large ionised bubbles around them, concluding that only after a very long period of continuous star formation, and assuming rather large values of the ionising continuum escape fraction, the ionised bubbles would reach such a size that Ly$\alpha$ photons would be able to escape the region unaffected by scattering in the intergalactic medium (IGM). They considered different options to explain the leakage of Ly$\alpha$ photons from only the three bright emitters, assuming that the faint LBGs could be more evolved, or located in the outskirts of the overdense region, still surrounded by neutral gas. 

In this paper we re-examine the data in \citet{Castellano2018} to check whether the complete collection of sources in the two overdense regions of the BDF would be capable of reionising two large bubbles around them. To this end, we have added the ionising flux from all the sources in each of the two regions of the BDF, including a set of still undetected, yet expected, low  luminosity sources.  

As we will discuss later, cosmological evolutionary models require only very low values of the ionising continuum escape fraction to explain the complete reionisation of the universe by $z \sim 6$ (around 5\% to 10\% in average).  These low values of $f{_{\rm esc,LyC}}$ are compatible with low values of the Ly$\alpha$ photon escape fraction in the line of sight, as shown by \citet{Chisholm2018}. As a good example, there is the prototypical Lyman Break Galaxy analogue Haro~11, with  $f{_{\rm esc,LyC}} = 0.03$  and $f{_{\rm esc,Ly\alpha}} = 0.04$ values reported by \citet{Verhamme2017}. \citet{Chisholm2018} considered indeed that in scenarios with low extinction, both $f{_{\rm esc,Ly\alpha}}$ and $f{_{\rm esc,LyC}}$ should be intrinsically very similar, though the scattering of Ly$\alpha$ photons by neutral clouds yields a non-predictable behaviour of the Ly$\alpha$ emission (see for example \citet{Dijkstra2016}). 
If the $f{_{\rm esc,Ly\alpha}}$ remain low, in line with the expected $f{_{\rm esc,LyC}}$ range, EW(Ly${\alpha}$) values would be well below the detection limit reported by  \citet{Castellano2018} (around 30~\AA\, at $z \sim 7$). Furthermore, while the destruction, by resonant scattering, of Ly$\alpha$ photons is a complex process, that depends largely on both the geometry and kinematics of the neutral gas close to the star-forming regions, we think that the lack of Ly$\alpha$ emission from the LBGs does not prevent the leakage of ionising photons when the full solid angle is considered. Therefore, the LBG galaxies could contribute, in a significant way, to the reionisation of the IGM even if they do not show  Ly$\alpha$ emission along the line of sight. 

To estimate the size of the ionised bubbles in the BDF, we have considered the two pointings reported by \citet{Castellano2018}. Then, we have  derived the Ly${\alpha}$ escape fraction of the three LAEs using the calibration by \citet{Sobral2019}.  Knowing the $f{_{\rm esc,Ly\alpha}}$ and the Ly${\alpha}$ fluxes from these three bright galaxies, we have derived the number of intrinsic ionising continuum photons. As for the medium and low luminosity sources, their ionising fluxes have been derived from the UV continuum using our own evolutionary models. Finally, we have used the AMIGA model \citep{Salvador-Sole2017} to derive values of the expected ionising emissivity at z $\sim 7$, with which we have compared the ionising fluxes from the BDF.   

Section~\ref{BDF} reviews the BDF field and the different sources within it. In particular, we have done an estimation of the number of low luminosity sources, based on the surface density derived by \citet{Bouwens2015} at z $\sim 7 $. Section~\ref{phot} shows the derivation of the number of ionising continuum photons produced by the different sources in the region. Finally, in Section~\ref{minimum} we use the expected emissivity at $z = 7$ from the AMIGA model  \citep{Salvador-Sole2017}, and considering the volumes of both regions,  we compute the minimum number of continuum ionising photons required to ionise each  region and explore the possible fomation of an even larger reionised bubble enclosing most of the BDF. We finish with the Conclusions in Section~\ref{conclusions}.  All units are in concordance cosmology units, namely ($\Omega_{\Lambda}$ = 0.7, $\Omega_{M}$=0.3, and $H_0 = 70$ Km/s/Mpc). Magnitudes are given in the AB system \citep{Oke1983}. For cosmological calculations we have used the \textit{CosmoCal} webtool kindly made available by \citet{Wright2006}, and the \textit{Cosmological Calculator for a Flat Universe} built by Nick Gnedin at Fermilab  ({\tt https://home.fnal.gov/$\sim$gnedin/cc/}).

\section{Characterising the observed Bremer Deep Field (BDF)}
\label{BDF}

The fields observed by \citet{Vanzella2011} and \citet{Castellano2018} were relatively small, around $0.7 \times 0.7$ pMpc$^2$ each, much smaller than the full Bremer Deep Field (BDF), which extends over $2.4 \times 2.4$ pMpc$^2$. The observations consisted of two pointings around BDF\,521 and  BDF\,3299, respectively. In what follows we will analyse separately the population of sources in these two pointings. As indicated in \citet{Castellano2016} each specific pointing covered 3.94 and 3.82 squared arcmin, respectively. \citet{Castellano2018} indicated that most of the galaxies identified should be within the redshift range $z \sim  6.95-7.15$. 
We have derived the volumes assuming the difference in distances between these limiting redshifts. Making use of the Cosmocalc \citep{Wright2006}, and of the \textit{Cosmological Calculator for a Flat Universe} by N. Gnedin we arrive to the volumes of each of the two pointings, namely 1718\,cMpc$^3$ for the region containing BDF\,521 and 1660\,cMpc$^3$ for the region containing BDF\,3229.  

\begin{figure}
\includegraphics[width=\columnwidth]{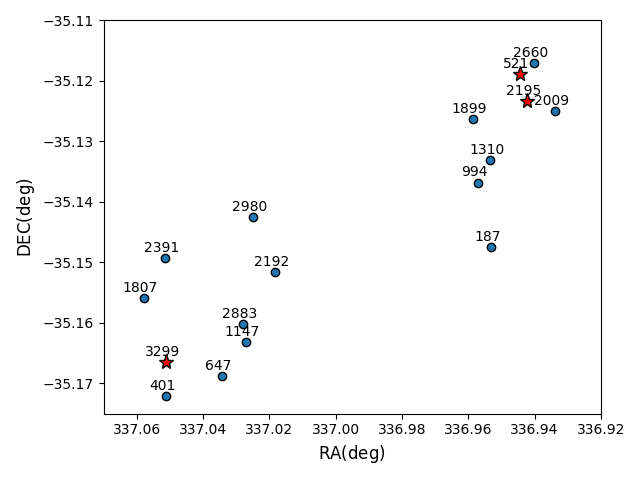} \\
    \caption{The two pointings can be noticed in this figure of the BDF galaxy distribution. The pointings were guided by the two LAEs discovered originally \citep{Vanzella2011}. The upper right one we will call Group\,1, while Group\,2 would be the bottom left one. The three LAEs are marked with solid red stars. The medium luminosity LBGs are marked with blue circles.}
    \label{fig:BDF_field}
\end{figure}

Within these two specific pointings \citet{Castellano2018} found 17 sources, three of which are Ly$\alpha$ emitters, while the rest are Lyman Break Galaxies with no Ly${\alpha}$ emission detected. For the three LAEs there is spectroscopy \citep{Castellano2018, Vanzella2011}, from which we have used both the Ly$\alpha$ fluxes and their equivalent widths. 

\begin{table*}
    \centering
    \begin{tabular}{c|c|c|c|c|c}
    \hline
    M$_{\rm AB}$ range & Surface Density & \#\, Group\,1 & \#\, Group\,2 & \#\,Corrected 1 & \#\,Corrected 2  \\
    & arcmin$^{-2}$ & 3.94\,arcmin$^2$ &3.82\,arcmin$^2$ & overdensity & overdensity\\
    \hline
     27.45 - 27.95 & $0.831 \pm 0.242$ & $3.27 \pm 0.92$ & 3.17 $\pm 0.95$ & $11.46 \pm 0.85$ &  $11.11 \pm 0.85$ \\
     27.95 - 28.45 & $1.273 \pm 0.300$ & $5.02 \pm 1.18$ & $4.86 \pm 1.15$ &  $17.55 \pm 4.38$ & $ 17.02 \pm 4.14$\\
     28.45 - 28.95 & $1.264 \pm 0.518$ & $4.98 \pm 2.04$ & $ 4.83 \pm 1.98 $ &  $17.43 \pm 7.14$ & $ 16.90 \pm 7.96$\\
     28.95 - 29.45 & $4.286 \pm 0.953$ & $16.37 \pm 3.64$ & $16.89 \pm 3.75$ & $57.10 \pm 13.14$ & $ 57.30 \pm 13.91$ \\
     29.45 - 29.95 & $3.484 \pm 0.859$ & $13.73 \pm 3.38$ & $13.31 \pm 3.28$ & $48.04 \pm 11.85$ & $ 46.58 \pm 12.53$\\
     \hline
    \end{tabular}
    \caption{Number of expected low luminosity sources in Group 1 and Group 2 of the BDF at $z \sim 7$. Columns 1) magnitude range, 2) surface density from \citet[Table A1]{Bouwens2015},  3) number of sources in Group\,1, 4) number of sources in Group\,2, 5) number of sources in Group\,1 multiplied by the overdensity factor, which we take as 3.5, and 6) number of sources in Group\,2 corrected as well by the same overdensity factor \citep{Castellano2016}.}
    \label{tab:lowL}
\end{table*}

\subsection{Number of low luminosity sources in the BDF}
\label{LowL}

Low luminosity galaxies are recognised as key elements in the process of re-ionising the Universe \citep{Bouwens2015, Robertson15, RodriguezEspinosa2020}. To derive the number of low luminosity sources expected in the BDF we have used the typical surface density values of high-redshift star-forming galaxies given in \citet{Bouwens2015} for the universe at $z \sim 7 $, as listed in their Appendix Table A1. The surface density and number of sources are included, for completeness, in Table~\ref{tab:lowL}. \citet{Castellano2016} also claim that the overdensity in the observed field ranges from 3 to 4. Thus, we  will assume an overdensity of 3.5 in what follows. Therefore, the derivation of the number of low luminosity sources has been done multiplying the surface density, in number of sources per arcmin$^2$, by the surface of the observed fields, 3.94~arcmin$^2$ and 3.82~arcmin$^2$, respectively. Then, the average number of low luminosity sources  in the BDF has been multiplied by 3.5, which is the average overdensity in the two small BD Fields that we have assumed according to \citet{Castellano2016}. The results are given in Table~\ref{tab:lowL}. Note that, the lowest luminosity in this case, m$_{AB}= 29.70$, corresponds to an absolute rest-frame UV magnitude of M$_{UV} = -19.49$.  It has been shown, that there are no cut offs in the UV luminosity function down to M$_{UV} \sim -15$ \citep{Bouwens2017,Yue2018}. Thus the results obtained herein for the low luminosity sources are rather conservative. Finally, we would like to mention that using the surface densities from \citet{Bouwens2015} we should expect to find barely one source brighter than 25.95. Note that, as expected, in the reduced BDF there is indeed only one source with m$_{AB}$ = 25.97, which is BDF2883.   

\begin{table*}
  \begin{tabular}{l|c|c|c|c|c|c|c}
       Name & Group & EW$_o$ & Flux$_{Ly\alpha}$ & $f_{\rm{esc,Ly}\alpha}$ & L$_{Ly\alpha}$ & L$_{Ly\alpha,intr}$& Q$_{ion,LAE}^{e}$\\
         &  & \rm{\AA} &  $10^{-17}$\,erg\,s$^{-1}$ && $10^{42}$\,erg\,s$^{-1}$ & $10^{43}$\,erg\,s$^{-1}$ & $10^{54}$s$^{-1}$ \\
         \hline
         BDF521 & 1 & $64 \pm 6$ & $1.62 \pm 0.16$& $0.32 \pm 0.03$ & $9.14 \pm 0.91$ & $2.98 \pm 0.41$ & $2.53 \pm 0.35$  \\
         BDF2195 & 1 & $50 \pm 12$ & $1.85 \pm 0.46$ & $0.24 \pm 0.06$ & $10.56 \pm 2.63$ & $4.40 \pm 1.52$ & $3.73 \pm 1.29$  \\
         BDF3299 & 2 & $50 \pm 6$ & $1.21 \pm 0.14$ &  $0.24 \pm 0.24$ & $7.08 \pm 0.83$ & $2.95 \pm 0.49$ & $2.50 \pm 0.42$  \\
         
    \hline
    \end{tabular}
    \caption{Lyman alpha emitters in the BDF. Name, Group, Ly$\alpha$ restframe equivalent width, Ly$\alpha$ flux, $f_{esc,Ly\alpha}$,  observed L$_{Ly\alpha}$ luminosity, intrinsic L$_{Ly\alpha}$ luminosity, effective  number of ionising continuum photons per second, Q$_{ion,LAE}^e$, respectively, of the three bright Lyman Alpha Emitting galaxies in \citet{Castellano2018}} 
    \label{tab:bright}
\end{table*}

\section{Number of ionising continuum photons from the sources in the BDF}
\label{phot}

To check whether the entire collection of sources reported in \citet{Castellano2018}, with the addition of the low luminosity sources that we have derived, is capable of producing two ionised bubbles or even a large one encompassing the whole region, we have first computed the number of ionising continuum photons 1) from the three LAEs, 2) from the rest of the galaxies reported by \citet{Castellano2018}, that we will call mid luminosity sources, and 3) from the expected low luminosity sources that we have derived from the field surface density at $z \sim 7$  \citep{Bouwens2015}.

 \begin{figure}
\includegraphics[width=\columnwidth]{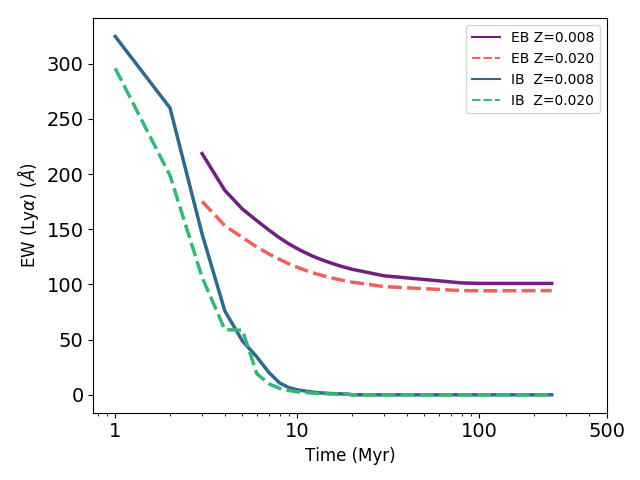}
\includegraphics[width=\columnwidth]{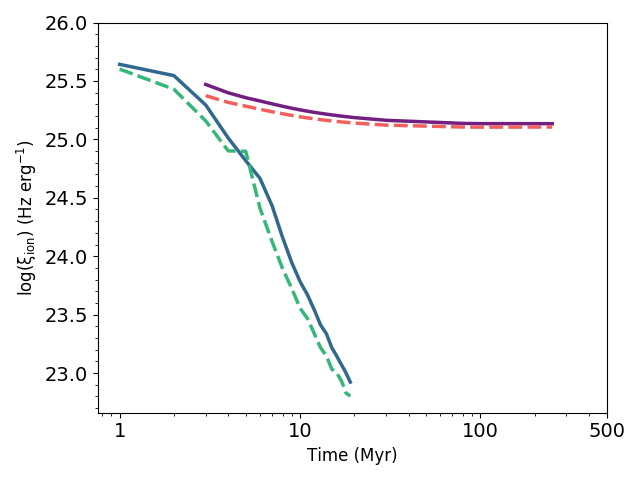} \\
    \caption{Ly$\alpha$ equivalent width (top) and \psion\ (bottom) evolution for instantaneous (IB) and extended (EB) episodes of star formation, for solar and subsolar metallicities, as predicted by \citet{Oti2010}. These models were based on the evolutionary code discussed by \citet{CMHK2002} for instantaneous bursts, and on Starburst~99 \citep{Leitherer1999} for extended episodes of star formation. In both cases they assumed a Salpeter Initial Mass Function with upper mass limit of 120~\modot.}
    \label{fig:EW}
\end{figure}

\subsection{Number of Lyman continuum photons from the three LAEs in the BDF}
\label{LAEs}

 There are three Ly$\alpha$ emitting galaxies in the BDF, two BDF521 and BDF3299, reported by \citet{Vanzella2011} and an additional  one, BDF 2195, identified by \citet{Castellano2018}. These three sources have redshifts derived from the spectroscopy, namely $z = 7.008$ for BDF521 and BDF2195 , and $z = 7.109$ for BDF3299 \citep{Castellano2018, Vanzella2011}. Moreover, for these sources we have also the measured rest-frame equivalent widths (EW$_{\rm o}$) \citep{Castellano2018}. To derive the number of ionising photons emitted per second from the observed Ly$\alpha$ fluxes  we need first to estimate the Ly$\alpha$ escape fraction, $f_{\rm esc,Ly\alpha}$. \citet{Sobral2019} derived a correlation between the Ly$\alpha$ equivalent width and the escape fraction of high-redshift galaxies, yielding an empirical relation of both parameters. Using that relation we have derived the values of $f_{\rm esc,Ly\alpha}$ for the 3 LAEs. From the escape fractions and the observed Ly$\alpha$ fluxes we get the  intrinsic Ly$\alpha$ luminosities.  Finally, we have computed the effective number of ionising photons per second (Q$^e_{\rm ion}$), i.e. the value corresponding to the intrinsic Ly$\alpha$ luminosities assuming case B conditions, using the expression $L(Ly\alpha) = 1.18\times10^{-11} \times Q^e_{ion}$ erg s$^{-1}$ \citep{osterbrock}. This relation does not depend on the properties of the ionising stars, nor on evolutionary models, but just on the physical conditions of the gas, and is based on a ratio $L(Ly\alpha)/L(H\alpha) = 8.7$ which is usually assumed for star-forming regions (see \citet{Dopita2003} and \citet{Hayes2019}). The results are listed in Table~\ref{tab:bright}.  Note that the intrinsic number of Lyman continuum photons (Q$_{\rm ion}^*$) being emitted by the massive stars will be larger by a factor $1/(1 - f_{\rm esc,LyC})$, since the escaping LyC photons do not participate in the ionisation of the gas traced by the Ly$\alpha$ emission. 
 
 We want to remark that the correlation by \citet{Sobral2019} implies that the intrinsic values of the Ly$\alpha$ equivalent widths converge in average around \EWlya\, $ \sim 200$~\AA. Indeed, most of the galaxies used to derive this correlation show intrinsic (once corrected from the escape fraction) \EWlya\ values within the range $150 - 250$ \AA. 
These high equivalent width values can only be achieved when stars with very high ionising power dominate the overall emission.  We show in Fig.~\ref{fig:EW}~(top) the predicted evolution of \EWlya\  by \citet{Oti2010} for a very short--lived starburst, and for a long--lasting episode forming massive stars at a constant rate during hundreds of Myr (a similar behaviour was presented by \citet{charlotfall1993} with a diferent set of models). Values of \EWlya\ above 200~\AA\ are predicted only during the first $\sim 3-4$~Myr after the onset of a massive star formation episode, but are not compatible with a starburst having formed stars at a stable rate during more than around 50~Myr, at least for metallicities above $Z \sim 0.008$. Most of the Ly$\alpha$ emitters at high redshift analysed by \citet{Sobral2019} should  therefore be experiencing very young massive star formation episodes, or a sudden, recent increase of their otherwise lower, long lasting star formation rate.

 The ionising power of a massive star cluster is generally defined in the literature as \psion = \Qint\ / L$_{1500}$, in units of \Herg \, (see \citet{MasHesse1991}) for an analysis of the evolution of the equivalent $B$ parameter as a function of the star formation scenario). Since we are dealing with Ly$\alpha$ equivalent widths in \AA, the following conversion applies:  \psion $\rm{(erg^{-1}}\rm{Hz)} = 1.35 \times 10^{23}$\, \EWlya\ (\AA). 
The average intrinsic value derived by \citet{Sobral2019}, \EWlya\, $ \sim 200~$\AA, corresponds to log (\psion) = 25.43~ \Herg .  We show in Fig.~\ref{fig:EW}~(bottom) the evolution of \psion\ for an instantaneous burst and an extended episode of star formation.

\subsection{The number of Lyman continuum photons from the medium Luminosity galaxies in the BDF}
\label{midL}

\begin{table*}
   \centering
    \begin{tabular}{l c c c c c c}
       Name & Group & M$_{\rm AB}$ & $f_{\lambda1310}$ & L$_{1500}$ & Q$_{\rm ion,LBG}^*$   \\
         & & &$10^{-20}$ erg\,s$^{-1}$\,cm$^{-2}$\,\AA$^{-1}$ & $10^{40}$ erg\,s$^{-1}$\,\AA$^{-1}$ & $10^{54}$~s$^{-1}$ \\
         \hline
         BDF2009 & 1 & 26.89$\pm0.08$ & 13.75$\pm12.78$ & 6.89$\pm5.61$ & 0.65$\pm0.50$   \\
         BDF994 & 1 & 27.11$\pm0.19$ & 11.23$\pm9.43$ & 5.69$\pm4.19$ &  0.53$\pm0.39$  \\
         BDF2660 & 1 & 27.27$\pm0.10$ & 9.69$\pm8.84$ & 4.91$\pm3.93$ &  0.46$\pm0.34$  \\
         BDF1310 & 1 & 27.32$\pm0.16$ & 9.26$\pm7.99$ & 4.69$\pm3.55$ & 0.44$\pm0.31$  \\
         BDF187 & 1 & 27.33$\pm0.10$ & 9.17$\pm8.37$ & 4.65$\pm3.72$ & 0.43$\pm0.33$  \\
         BDF1899 & 1 & 27.35$\pm0.15$ & 9.00$\pm7.84$ & 4.56$\pm3.49$ & 0.42$\pm0.31$  \\
         &&&& {\textbf{Total Group\,1}} & {\textbf{2.93$\pm$0.15} }\\
         \hline
         BDF2883 & 2 & 25.97$\pm$ 0.08 & 32.10$\pm29.82$ & 16.08$\pm14.94$ & 1.51$\pm1.32$   \\
         BDF401 & 2 & 26.43$\pm$ 0.08 & 21.01$\pm19.52$ & 10.53$\pm9.78$ & 0.99$\pm0.87$  \\
         BDF1147 & 2 & 27.26$\pm 0.11$ & 9.78$\pm8.84$& 4.72$\pm4.23$ &  0.46$\pm0.39$  \\
         BDF2980 & 2 & 27.30$\pm 0.12$ & 9.43$\pm8.44$ &  4.72$\pm4.23$ & 0.44$\pm0.38$  \\
         BDF647 & 2 & 27.31$\pm0.15$ & 9.34$\pm8.14$ & 4.68$\pm4.08$ & 0.44$\pm0.36$  \\
         BDF2391 & 2 & 27.33$\pm0.17$ & 9.17$\pm7.84$ & 4.59$\pm3.93$ & 0.43$\pm0.35$  \\
         BDF1807 & 2 & 27.36$\pm0.09$ & 8.92$\pm8.21$ & 4.47$\pm4.11$ & 0.42$\pm0.36$ \\
         BDF2192 & 2 & 27.40$\pm0.10$ & 8.60$\pm 7.84$ & 4.31$\pm3.93$ & 0.41$\pm0.35$  \\
          &&&& {\textbf{Total Group\,2}}&  {\textbf{5.11$\pm$0.64}} \\
          \hline
    \end{tabular}
    \caption{Lyman Break Galaxies in the BDF with no Ly$\alpha$ emission detected. 1) Name, 2) group, 3) observed magnitude M$_{AB}$, flux at $\lambda1310$ rest-frame, 4) luminosity at $\lambda1500$ and 5) number of intrinsic Lyman continuum photons emitted per second assuming an extended episode of star formation \citep{Oti2010}. }
    \label{tab:all}
\end{table*}

To derive the number of ionising continuum photons generated by the Lyman Break Galaxies in the BDF, we have computed first their luminosities in the rest UV band. We expect that all the sources are within a redshift range $ 6.95 <  z < 7.15$. This assumes that the galaxies in the observed BDF are part of the same structure \citep{Castellano2018}.  Indeed, as there is no spectroscopic confirmation, precise redshifts of the sources are not known. 
For convenience, we will assume the central wavelength of the Y105 filter, used for their discovery, as corresponding to the continuum at rest 1310~\AA,  at the redshift of 7.008. We have derived the $f_{1310}$ directly from the AB magnitudes given by \citet{Castellano2018}  correcting the fluxes to the rest-frame wavelength, and computing $L_{1310}$ assuming that $z = 7.008$ is valid for all galaxies. 
We have assumed that $L_{1500} = 0.88 \times L_{1310}$ (corresponding to the mean UV continuum slope expected from a population of young, massive stars, with no extinction), and have used the predictions from the evolutionary models of \citet{Oti2010}, as available in their webtool\footnote{\tt http://sfr.cab.inta-csic.es/index.php}, to estimate the number of Lyman continuum photons being produced by the stars, for each value of the UV continuum luminosity. This webtool allows to estimate the intrinsic number of continuum ionising photons emitted by the starburst as a function of various parameters, including $L_{1500}$, and for different star formation scenarios. 

Since the LBGs have no Ly$\alpha$ emission detected, we do not have any hint that these galaxies could be experiencing a very young star formation episode. If we assume these galaxies are in any case experiencing a recent, or still active, episode of massive star formation, their intrinsic \EWlya\ (or \psion) values should be in between the predictions for the two scenarios considered in Fig.~\ref{fig:EW}: a very young instantaneous burst or an extended episode having already reached an equilibrium between the birth an death of the most massive stars, i.e., active during more than around the last 50\~Myr. After this time,  the \EWlya\~ evolves very slowly and can be considered constant for up to around 1\~Gyr, longer than the age of the universe at $z \sim 7$ (see also the predictions by \citet{charlotfall1993}). \citet{Oti2010} predict that the \Lya\  equivalent width would converge to \EWlya\ $\sim 95 - 101$ \AA\ (or log(\psion) $\sim 25.10 - 25.13$ \Herg)   for metallicities in the range $Z = 0.020 - 0.008$. The weak dependence on metallicity is related to the fact that as the massive stellar population stabilises with time, both the ionising and the UV continuum flux of low metallicity stars increase when compared to solar metallicity, so that the effect on the \EWlya\ is partially compensated.

We have therefore assumed as our initial scenario the more conservative option of an extended star forming process having reached the equilibrium phase.   In Table~\ref{tab:all} we list the derived luminosities at rest 1500~\AA, $L_{1500}$, and the corresponding number of ionising continuum photons (Q$^*_{ion}$) value, as estimated with the webtool for extended episodes at 250~Myr, assuming an intrinsic \EWlya\ $\sim 100$ \AA. Note that we have corrected \Qint\ from the standard 30\% destruction factor assumed by the \citet{Oti2010} models, not applicable to high-redshift, low dust galaxies.  We insist that since we are dealing with equivalent widths, which are tracing the ionising power per UV luminosity unit, our estimates are essentially independent on the precise history of star formation, but are only scaled to the UV continuum luminosity. The same \Qint\ values would be derived for star formation episodes with stable rates during the last 50 to 500 Myr. On the other hand, the predictions for an instantaneous burst should provide the expected upper limit if the star formation rate has suffered a very recent, significant increase in the last $1 - 5$ Myr. 

The assumed intrinsic \Lya\  value (\EWlya\ $\sim 100$ \AA, or log \psion\  $\sim 25.13$ \Herg) is very similar to the canonical values of \citet{Kennicut1998}, log \psion\  = 25.11\, \Herg\ (see \citet{Bouwens2016}).  We want to stress that most of these values are well constrained by the predictions of our evolutionary models as shown in Fig.~\ref{fig:EW}.

\begin{table}
    \centering
    \begin{tabular}{c|c|c|c|c}
    m$_{\rm AB}$ & $f_{\lambda1310}$ & L$_{1500}$ & Q$_{\rm ion,G1}^*$ & Q$_{\rm ion,G2}^*$ \\
    & $\times 10^{-20}$ & $ \times 10^{40}$ & $\times 10^{54}$ & $\times 10^{54}$ \\
    & ${\rm erg}$\,s$^{-1}$\,cm$^{-2}$\,\AA$^{-1}$ & ${\rm erg}$\,s$^{-1}$ & s$^{-1}$ & s$^{-1}$\\
\hline
    27.70 & 6.74 & 3.38 & 3.64$\pm 0.38$ & 3.53$\pm 0.27$  \\
    28.20 & 4.25 & 2.13 & 3.52$\pm 0.30$ & 3.41$\pm 0.21$ \\
    28.70 & 2.68 & 1.34 & 2.20$\pm 0.33$ & 2.14$\pm 0.23$ \\
    29.20 & 1.69 & 0.85 & 4.72$\pm 0.38$ & 4.57$\pm 0.27$ \\
    29.70 & 1.07 & 0.54 & 2.42$\pm 0.22$ & 2.35$\pm 0.15$  \\
\hline
 Total  &&& 16.49$\pm 0.39$ & 15.99$\pm 0.27$ \\
    \hline
    \end{tabular}
    \caption{Non-detected low luminosity sources. Columns 1) average magnitude AB, 2) flux at 1310\AA, 3) continuum luminosity at 1500\AA, 4) number of ionising continuum photons, once corrected for the number of sources and overdensity for Group\,1 and 5) for Group\,2.}
    \label{tab:lowLQ}
\end{table}

\subsection{The output from the low luminosity sources}
\label{lowL}

Likewise, we have derived the number of Lyman continuum photons for the expected, non-detected, lower luminosity sources, which play an essential role in producing the ionising photons required to re-ionise the Universe \citep{Bouwens2015,Higuchi2019,Tilvi2020}. First, we have computed the luminosity at 1500~\AA\, corresponding to the middle AB magnitude for each of the magnitude ranges listed in Table~\ref{tab:lowL}, following the same procedure as in the previous section. Thus, using a similar methodology as in the case of the medium luminosity galaxies, we have derived the expected number of ionising continuum photons corresponding to each AB magnitude bin, considering the total number of expected galaxies in each magnitude range as listed in Table~\ref{tab:lowL}. We list in Table~\ref{tab:lowLQ} the derived number of intrinsic ionising continuum photons expected from the
low luminosity galaxies. When compared with the contributions by the LAEs and LBGs in the sample, it becomes evident that faint galaxies indeed dominate the release of ionising photons to the IGM, as already proposed by \citet{Bouwens2016}, \citet{Castellano2018} and \citet{Robertson2013}. We would like finally to remind that these estimates should be considered as lower limits, since we have assumed extended star formation episodes in their equilibrium phase, as for the LBGs in the previous section, with canonical values for the ionising photon production efficiency. We'll discuss in Sect.~\ref{bubble} the effect of other possible scenarios.

\begin{table}
    \centering
    \begin{tabular}{c|l|l}
    
    & Q$_{\rm ion,G1}^*$ & Q$_{\rm ion,G2}^*$ \\
    & $\times10^{54}\,{\rm s}^{-1}$ & $\times10^{54}\,{\rm s}^{-1}$ \\
    \hline
    LAEs & $\gtrsim 6.26\pm 0.94$ & $\gtrsim 2.50 \pm 0.42$\\
    Mid Luminosity & $2.93\pm0.15$ & $5.11\pm0.64$\\
    Low Luminosity & $16.49\pm 0.39$ & $15.99\pm 0.27$ \\
    \hline
    Total & $\leq 25.68 \pm 0.59$ & $ \leq 23.60 \pm 0.47$ \\
    \hline
    \end{tabular}
    \caption{The number of intrinsic ionising continuum photons produced by the medium and low luminosity LBGs in each of the Groups. The values listed for the LAEs are just lower limits as they should be corrected by the corresponding Lyman continuum escape fraction as discussed in the text.}
    \label{tab:sum}
\end{table}

\section{Discussion}
\label{minimum}

A non-zero Lyman continuum escape fraction implies that the escaping photons will be able to ionise regions farther out from the galaxy or galaxies that contain the massive stars producing the ionising photons. Adding to the number of ionising continuum photons produced by the LAEs the contribution by the medium and low luminosity star forming galaxies for both Group\,1 and Group\,2 in the BDF,  we arrive to a total number of ionising continuum photons of $\gtrsim 25.68 \pm 0.59 \times 10^{54}$ ~s$^{-1}$, for Group\,1 and $\gtrsim 23.60 \pm 0.47 \times 10^{54}$ ~s$^{-1}$ for Group\,2, as listed in Table~\ref{tab:sum}. We remark that for the LAEs we  don't yet know the intrinsic number of ionising continuum photons emitted per second, since it depends on the actual LyC escape fraction. 

The next step is to derive the  minimum number of ionising continuum photons necessary to fully ionise the volumes of both Group\,1 and Group\,2. We get these values by multiplying the ionising emissivities derived from the AMIGA model \citep{Salvador-Sole2017} by the volumes of Group\,1 and Group\,2. AMIGA, the {\em  Analytic Model of Intergalactic-medium and Galaxies},  is a very complete and detailed self-consistent model of galaxy formation, particularly well suited to monitor the intertwined evolution of both luminous sources and the IGM. It computes the instantaneous emission at all the relevant wavelengths of normal galaxies, including their intrinsic ionising power, using the evolutionary synthesis models by \citet{Bruzual2003}, assuming a Salpeter Initial Mass Function. The parameters in the AMIGA model have been tuned to reproduce as well as possible the properties of high redshift star forming galaxies, including the contribution of very low metallicity Pop~III stars in the early epochs. Therefore, the average ionising emissivities derived from AMIGA should provide a fair estimate of the number of ionising continuum photons necessary to ionise the IGM at $z \sim 7$. AMIGA distinguishes between the most usual case of single reionisation, plus the less usual case of double reionisation. These emissivities, as derived from AMIGA, are 0.26 $\pm$ $0.01 \times 10^{51}$ {\rm s}$^{-1}{\rm cMpc}^{-3}$ for the single reionisation scenario, and 0.36 $\pm$ $0.01 \times 10^{51}${\rm s}$^{-1}{\rm cMpc}^{-3}$ for the double reionisation case. 

Though the AMIGA simulations include already the presence of clumpiness, the predicted emissivities correspond to average values over the IGM at $z \sim 7$. Since the two regions we are considering in the BDF seem to be overdense by a factor 3 -- 4 \citep{Castellano2016}, we consider that the density of ionising photons should also be larger by a factor $\sim3.5$ to fully reionise the local IGM. We have therefore multiplied the emissivities derived from AMIGA by 3.5 to properly take this effect into account. Moreover, since the average ionised fraction of the IGM predicted by AMIGA  at z $\sim 7$ is only $\sim 0.7$ for both the single and the double reionisation scenarios, we have divided the emissivities by this value to account for a fully reionised IGM. 

The emissivities after these corrections are 1.28 $\pm$ $0.05 \times 10^{51}$ {\rm s}$^{-1}{\rm cMpc}^{-3}$ for single reionisation and 1.75$\pm$ $0.05 \times 10^{51}$ {\rm s}$^{-1}{\rm cMpc}^{-3}$ for double reionisation. Finally, multiplying these emissivities by the volumes of each of the Groups we derive the minimum number of continuum ionising photons per second required to completely ionise those volumes. The resulting values are 2.20 $\pm$ $0.09 \times 10^{54}$ {\rm s}$^{-1}$ for Group\,1 and 2.12 $\pm$ $0.09  \times 10^{54}$ {\rm s}$^{-1}$ for Group\,2 in the case of single reionisation. For the case of double reionisation the values are 3.01 $\pm$ $0.09 \times 10^{54}$ {\rm s}$^{-1}$ for Group\,1 and 2.91 $\pm$ $0.08 \times 10^{54}$ {\rm s}$^{-1}$ for Group\,2, as listed in Table~\ref{tab:fesc}. 

The total number of photons available for reionising the circumgalactic medium (CGM) will depend on the Lyman continuum escape fraction, $f_{\rm esc,LyC}$. We can constrain the average value required to fully reionise the volumes around Group~1 and Group~2 by comparing the yield of ionising photons we have derived with the predictions by AMIGA, i.e., by solving equation~\ref{eq1}:

\begin{equation}
   \left\{{Q_{\rm ion}^*} + \frac{Q_{ion,LAE}^e}{1-f_{\rm esc,LyC}} \right\}\times {f_{\rm esc,LyC}} = \dot {\rm N}_{min,corr}
   \label{eq1}
\end{equation}

The first term of equation~\ref{eq1} corresponds to the intrinsic number of ionising continuum photons produced by the massive stars in the medium and low luminosity galaxies. The second term concerns the LAEs.  In the case of the LAES, to derive the intrinsic number of Lyman continuum photons we have to divide the {\em effective} number  by $\frac{1}{1 - f_{\rm esc,LyC}}$. Then, we multiply these two terms by the Lyman continuum escape fraction to get the number of continuum ionising photons available to ionise the IGM. Finally, the term at the other side of equation~\ref{eq1} is the emissivity derived from the AMIGA model, multiplied by the volumes of each Group, and corrected by both the overdensity and ionisation fraction at $z \sim 7$.

\begin{table}
    \centering
    \begin{tabular}{l|c|c|c|c}
      &${\dot N}_{min,corr}^{S}$ & ${\dot N}_{min,corr}^{D}$ &${\rm f}_ {esc,S} $&${\rm f}_ {esc,D}$  \\
      & $\times 10^{54}$ s$^{-1}$ & $\times 10^{54}$ s$^{-1}$ & &\\
    \hline
     Group\,1 & 2.20 $\pm$ 0.09  & 3.01 $\pm$ 0.09  & 0.08 & 0.10\\
     Group\,2 
     & 2.12 $\pm$ 0.08 & 2.91 $\pm$ 0.08 & $0.12$ & $0.14 $ \\
    \hline
    \end{tabular}
    \caption{Lyman continuum escape fractions derived for Group\,1 and Group\,2 in the single and double reionisation scenarios.These escape fraction are enough to fully ionise both regions. }
    \label{tab:fesc}
\end{table}

The results, as shown in Table~\ref{tab:fesc}, indicate that the volumes of Group\,1 and Group\,2 in the BDF would be completely ionised if the escape fractions of Lyman continuum photons are as low as 0.08 for Group 1 and 0.12 for Group 2, in the case of single reionisation. For the case of double reionisation the two groups would be fully ionised if the $f_{\rm esc,LyC}$ are 0.10 and 0.14 respectively. These are rather low values of the average Lyman continuum escape fractions. Our method allows to constrain the average $f_{\rm esc,LyC}$ value, but we want to stress that values for specific galaxies can vary significantly. \citet{Finkelstein2019} proposed that $f_{\rm esc,LyC}$ should be inversely correlated with the halo mass of the individual galaxies, thus $f_{\rm esc,LyC}$ becoming significantly higher for galaxies with $M_h < 10^{7.5}$ \modot\ (see their Figure~2). Since the main contributors to the ionising power in the BDF are the low luminosity galaxies, we interpret our results in the sense that the derived $f_{\rm esc,LyC}$ values should represent the typical values for the low mass galaxies. 

Such low values of $f_{\rm esc,LyC}$ are consistent with the fact that there are only three LAEs in the BDF. According to \citet{Chisholm2018} we should expect similar low values of $f_{\rm esc,LyC}$ and $f_{\rm esc,Ly\alpha}$ in galaxies with low extinction by dust, as should be the case at $z \sim 7$ \citep{Hayes2011}. With an average $f_{\rm esc,LyC} \sim 0.1$, and following \citep{Chisholm2018} and the calibration by \cite{Sobral2019} we would expect EW(Ly$\alpha$) $\sim 20$~\AA, which is too low to be detected on the BDF observations. Nevertheless, the large dispersion expected in $f_{\rm esc,Ly\alpha}$ \citep{Dijkstra2016} would support the presence of some galaxies with larger values of the Ly$\alpha$ escape fraction, which would come out as the the three LAEs  identified in the BDF. 

\subsection{A large ionised bubble in the BDF}
\label{bubble}

We want to stress that our analysis has been rather conservative when deriving the number of intrinsic ionising photons emitted by the galaxies in the BDF. The calibration of \fesclya\ vs. the rest--frame \EWlya\ by \citet{Sobral2019} for the LAEs implies an average intrinsic value of the \Lya\ equivalent around $\sim 200$ \AA, corresponding to log \psion\ $\sim 25.43$ \Herg, while \citet{Harikane2018} finds an average log \psion\  $\sim 25.53$ \Herg\, (and \fesclyc\ $\sim 0.10$) for a  large sample of LAEs at $z \sim  4.9 - 7.0$. On the other hand, as discussed above, the assumed intrinsic \Lya\  value \EWlya\ $\sim 100$ \AA\ for the LBGs corresponds to log  \psion $\sim 25.13$ \Herg. \citet{Bouwens2016} derived average values log \psion\  $\sim  25.3 $ \Herg\ for a sample of galaxies with \textit{Spitzer} H$\alpha$ IRAC observations  at $z \sim 4 - 5$, with an intrinsic scatter of $\sim 0.3$ dex for individual galaxies. The UV continuum bluest galaxies in the sample reached log \psion\ $\sim  25.5 - 25.8$  \Herg , indicating that \fesclyc\ cannot be in excess of 0.13. \citet{Stark2015} derived log  \psion\ $ = 25.68$ \Herg\ from the CIV $\lambda1548$ observations of A1703-zd6, a galaxy at $z = 7.045$ with \Lya\  EW$_o$ $\sim 65$ \AA\ \citep{Schenker2012}, and \citet{Stark2017} derived Log \psion\  $\sim 25.58$ \Herg\,   for three luminous ($M_{UV} = -22$) galaxies at $z = 7.15$, 7.48 and 7.73. Moreover, \citet{Bouwens2015} estimate log \psion\ =  $25.46$ \Herg\,  for faint galaxies at $z \sim  7 - 8$, with an associated $f_{\rm esc,LyC} \sim 0.11$, to properly match the reionisation timeline of the Universe. Finally, \citet{Wilkins2016} constrained  log \psion\ during the reionisation epoch to the range $25.1 - 25.5$ \Herg\, by combining the BlueTides cosmological hydrodynamical simulation with a range of stellar population synthesis models.

Large values of the intrinsic \EWlya\ or \psion\ can be associated to very young star formation episodes (or to a sudden increase of the star formation rate in the last $\sim 10$~Myr), or either to the presence of very low metallicity stars. \citet{schaerer2003} showed that for metallicities down to $Z \sim 10^{-7}$ the intrinsic values of \EWlya\ can reach up to \EWlya\ $ \sim 250 - 300$ \AA\ for stable star formation rates in the equilibrium phase, and even higher values for lower metallicities and/or larger values of the Initial Mass Function upper mass limit. Combining all these observational results and model predictions, we consider that the intrinsic \EWlya\ values of the LAEs, LBGs and low luminosity galaxies we have used in our calculations could realistically be increased by a factor 4 at most. Keeping \fesclyc\ $ \sim 0.1$, as constrained by the results discussed above, this would lead to a total number of ionising continuum photons released to the IGM of $ 1.05 \times 10^{55}$ ~s$^{-1}$, for Group\,1 and $ 0.96 \times 10^{55}$ ~s$^{-1}$ for Group\,2, i.e., a total \Qint\ $\sim 2 \times 10^{55}$ ~s$^{-1}$. 

Comparing these numbers with the AMIGA emissivities listed in Table~\ref{tab:fesc} we conclude that an IGM volume 4.6 times (single reionisation) or 3.4 times (for double reionisation)  larger than the volumes of Group~1 and Group~2 together, would become reionised assuming larger, but still realistic, values of the ionising power for the star--forming galaxies in the two BDF overdensities. Since the volume comprising both overdensities (extended over 30~arcmin$^2$) would be roughly a factor $\sim 4$ larger than the added volume of both groups, there would be enough ionising photons released to the IGM for the two ionised bubbles around them to merge in a single, very large re-ionised bubble.

We conclude that realistic, rather small values of the Lyman continuum escape fractions would allow to completely reionise the overdense regions of the BDF within a single, large ionised bubble, such as those that through percolation completed the reionisation of the universe by $z \sim 6$. On the other hand, the still required presence of neutral gas around the star forming regions in these galaxies, evidenced by the low values of the escape fractions, would explain the scarcity of detected faint Ly$\alpha$ emitters, since the observed EW(Ly$\alpha$) values would remain, on average, rather low.

\section{Conclusions}
\label{conclusions}

 We have looked for the reionisation status of the Intergalactic Medium around two overdense groups of star--forming galaxies in the Bremer Deep Field. To this end, we have considered all the sources in the BDF, including galaxies that are expected but have not yet been detected. These are  low luminosity sources, for which we have estimated their numbers in the BDF assuming the average surface density at $z \sim 7$ from \citet{Bouwens2015} and an overdensity of $\times 3.5$, as estimated by \citet{Castellano2016}. Then we have derived the number of intrinsic ionising photons from the bright \Lya\ emitters and the medium luminosity Lyman Break Galaxies identified by \citet{Castellano2018}, to which we have added the contribution of the expected low luminosity sources.  Even adopting conservative estimates for the ionising continuum photons produced by the massive stars in all these sources, known and expected, we conclude that there would be enough photons to ionise two large bubbles, one per Group in the BDF, with average LyC escape fraction as low as $f_{\rm esc,LyC} \sim 0.08$ for Group\,1 and 0.12 for Group\,2. With less conservative, but more realistic, estimates of the ionising power, the two bubbles would be merging into a large ionised bubble comprising most of the galaxies in the BDF. These low values of the LyC escape fraction indicate that there are still substantial amounts of neutral hydrogen surrounding the star forming regions in these galaxies. The inferred low values of the $f_{\rm esc,Ly\alpha}$ would explain the scarcity of  faint Ly$\alpha$ emitters found within the nonetheless completely reionised bubbles. We confirm previous hints indicating  that the low luminosity sources are indeed the ones that dominate the reionisation of the BDF. Finally, we note that a scenario with a double reionisation would require only slightly larger LyC escape fractions than the more commonly assumed single reionisation.
 
\bigskip
\noindent \textbf{Data Availability}

All the LAEs Fluxes and restframe \EWlya\ values, as well as the medium Luminosity LBG magnitudes are available in \citet{Castellano2018}. The low luminosity sources data are new in this paper. 

\bigskip
\noindent \textbf{Acknowledgements}

We are very grateful to Sonia Torrej\'on de Pablos for having computed the \EWlya\ predictions plotted in Figure~\ref{fig:EW}. We are also very grateful to Dr. Alberto Manrique (U. of Barcelona) for sharing his values of the AMIGA emissivities. JMRE acknowledges the Spanish State Research Agency under grant number AYA2017-84061-P and is indebted to the Severo Ochoa Programme at the IAC. JMMH is funded by Spanish State Research Agency grants PID2019-107061GB-C61 and MDM-2017-0737 (Unidad de Excelencia Mar\'{\i}a de Maeztu CAB).  

\bibliographystyle{mnras}
\bibliography{TheBDF}

\begin{thebibliography}{}
\makeatletter
\relax
\def\mn@urlcharsother{\let\do\@makeother \do\$\do\&\do\#\do\^\do\_\do\%\do\~}
\def\mn@doi{\begingroup\mn@urlcharsother \@ifnextchar [ {\mn@doi@}
  {\mn@doi@[]}}
\def\mn@doi@[#1]#2{\def\@tempa{#1}\ifx\@tempa\@empty \href
  {http://dx.doi.org/#2} {doi:#2}\else \href {http://dx.doi.org/#2} {#1}\fi
  \endgroup}
\def\mn@eprint#1#2{\mn@eprint@#1:#2::\@nil}
\def\mn@eprint@arXiv#1{\href {http://arxiv.org/abs/#1} {{\tt arXiv:#1}}}
\def\mn@eprint@dblp#1{\href {http://dblp.uni-trier.de/rec/bibtex/#1.xml}
  {dblp:#1}}
\def\mn@eprint@#1:#2:#3:#4\@nil{\def\@tempa {#1}\def\@tempb {#2}\def\@tempc
  {#3}\ifx \@tempc \@empty \let \@tempc \@tempb \let \@tempb \@tempa \fi \ifx
  \@tempb \@empty \def\@tempb {arXiv}\fi \@ifundefined
  {mn@eprint@\@tempb}{\@tempb:\@tempc}{\expandafter \expandafter \csname
  mn@eprint@\@tempb\endcsname \expandafter{\@tempc}}}

\bibitem[\protect\citeauthoryear{{Abdullah}, {Wilson}  \& {Klypin}}{{Abdullah}
  et~al.}{2018}]{Abdullah2018}
{Abdullah} M.~H.,  {Wilson} G.,   {Klypin} A.,  2018, \mn@doi [\apj]
  {10.3847/1538-4357/aac5db}, \href
  {https://ui.adsabs.harvard.edu/abs/2018ApJ...861...22A} {861, 22}

\bibitem[\protect\citeauthoryear{{Bouwens}, {Illingworth}, {Blakeslee}  \&
  {Franx}}{{Bouwens} et~al.}{2006}]{Bouwens2006}
{Bouwens} R.~J.,  {Illingworth} G.~D.,  {Blakeslee} J.~P.,   {Franx} M.,  2006,
  \mn@doi [\apj] {10.1086/498733}, \href
  {http://adsabs.harvard.edu/abs/2006ApJ...653...53B} {653, 53}

\bibitem[\protect\citeauthoryear{{Bouwens} et~al.,}{{Bouwens}
  et~al.}{2010}]{Bouwens2010}
{Bouwens} R.~J.,  et~al., 2010, \mn@doi [\apj] {10.1088/0004-637X/725/2/1587},
  \href {http://adsabs.harvard.edu/abs/2010ApJ...725.1587B} {725, 1587}

\bibitem[\protect\citeauthoryear{{Bouwens} et~al.,}{{Bouwens}
  et~al.}{2015}]{Bouwens2015}
{Bouwens} R.~J.,  et~al., 2015, \mn@doi [\apj] {10.1088/0004-637X/803/1/34},
  \href {https://ui.adsabs.harvard.edu/abs/2015ApJ...803...34B} {803, 34}

\bibitem[\protect\citeauthoryear{{Bouwens}, {Smit}, {Labb{\'e}}, {Franx},
  {Caruana}, {Oesch}, {Stefanon}  \& {Rasappu}}{{Bouwens}
  et~al.}{2016}]{Bouwens2016}
{Bouwens} R.~J.,  {Smit} R.,  {Labb{\'e}} I.,  {Franx} M.,  {Caruana} J.,
  {Oesch} P.,  {Stefanon} M.,   {Rasappu} N.,  2016, \mn@doi [\apj]
  {10.3847/0004-637X/831/2/176}, \href
  {https://ui.adsabs.harvard.edu/abs/2016ApJ...831..176B} {831, 176}

\bibitem[\protect\citeauthoryear{{Bouwens}, {Oesch}, {Illingworth}, {Ellis}  \&
  {Stefanon}}{{Bouwens} et~al.}{2017}]{Bouwens2017}
{Bouwens} R.~J.,  {Oesch} P.~A.,  {Illingworth} G.~D.,  {Ellis} R.~S.,
  {Stefanon} M.,  2017, \mn@doi [\apj] {10.3847/1538-4357/aa70a4}, \href
  {https://ui.adsabs.harvard.edu/abs/2017ApJ...843..129B} {843, 129}

\bibitem[\protect\citeauthoryear{{Bruzual} \& {Charlot}}{{Bruzual} \&
  {Charlot}}{2003}]{Bruzual2003}
{Bruzual} G.,  {Charlot} S.,  2003, \mn@doi [\mnras]
  {10.1046/j.1365-8711.2003.06897.x}, \href
  {https://ui.adsabs.harvard.edu/abs/2003MNRAS.344.1000B} {344, 1000}

\bibitem[\protect\citeauthoryear{{Calvi} et~al.,}{{Calvi}
  et~al.}{2019}]{Calvi2019}
{Calvi} R.,  et~al., 2019, \mn@doi [\mnras] {10.1093/mnras/stz2177}, \href
  {https://ui.adsabs.harvard.edu/abs/2019MNRAS.489.3294C} {489, 3294}

\bibitem[\protect\citeauthoryear{{Castellano} et~al.,}{{Castellano}
  et~al.}{2016}]{Castellano2016}
{Castellano} M.,  et~al., 2016, \mn@doi [\apjl] {10.3847/2041-8205/818/1/L3},
  \href {https://ui.adsabs.harvard.edu/abs/2016ApJ...818L...3C} {818, L3}

\bibitem[\protect\citeauthoryear{{Castellano} et~al.,}{{Castellano}
  et~al.}{2018}]{Castellano2018}
{Castellano} M.,  et~al., 2018, \mn@doi [\apjl] {10.3847/2041-8213/aad59b},
  \href {https://ui.adsabs.harvard.edu/abs/2018ApJ...863L...3C} {863, L3}

\bibitem[\protect\citeauthoryear{{Cervi{\~n}o}, {Mas-Hesse}  \&
  {Kunth}}{{Cervi{\~n}o} et~al.}{2002}]{CMHK2002}
{Cervi{\~n}o} M.,  {Mas-Hesse} J.~M.,   {Kunth} D.,  2002, \mn@doi [\aap]
  {10.1051/0004-6361:20020785}, \href
  {https://ui.adsabs.harvard.edu/abs/2002A&A...392...19C} {392, 19}

\bibitem[\protect\citeauthoryear{{Chanchaiworawit} et~al.,}{{Chanchaiworawit}
  et~al.}{2017}]{Kritt2017}
{Chanchaiworawit} K.,  et~al., 2017, \mn@doi [\mnras] {10.1093/mnras/stx782},
  \href {http://adsabs.harvard.edu/abs/2017MNRAS.469.2646C} {469, 2646}

\bibitem[\protect\citeauthoryear{{Chanchaiworawit} et~al.,}{{Chanchaiworawit}
  et~al.}{2019}]{Chanchaiworawit2018}
{Chanchaiworawit} K.,  et~al., 2019, \mn@doi [\apj] {10.3847/1538-4357/ab1a34},
  \href {https://ui.adsabs.harvard.edu/abs/2019ApJ...877...51C} {877, 51}

\bibitem[\protect\citeauthoryear{{Charlot} \& {Fall}}{{Charlot} \&
  {Fall}}{1993}]{charlotfall1993}
{Charlot} S.,  {Fall} S.~M.,  1993, \mn@doi [\apj] {10.1086/173187}, \href
  {https://ui.adsabs.harvard.edu/abs/1993ApJ...415..580C} {415, 580}

\bibitem[\protect\citeauthoryear{{Chiang}, {Overzier}, {Gebhardt}  \&
  {Henriques}}{{Chiang} et~al.}{2017}]{Chiang2017}
{Chiang} Y.-K.,  {Overzier} R.~A.,  {Gebhardt} K.,   {Henriques} B.,  2017,
  \mn@doi [\apjl] {10.3847/2041-8213/aa7e7b}, \href
  {https://ui.adsabs.harvard.edu/abs/2017ApJ...844L..23C} {844, L23}

\bibitem[\protect\citeauthoryear{{Chisholm} et~al.,}{{Chisholm}
  et~al.}{2018}]{Chisholm2018}
{Chisholm} J.,  et~al., 2018, \mn@doi [\aap] {10.1051/0004-6361/201832758},
  \href {https://ui.adsabs.harvard.edu/abs/2018A&A...616A..30C} {616, A30}

\bibitem[\protect\citeauthoryear{{Dijkstra}, {Gronke}  \&
  {Venkatesan}}{{Dijkstra} et~al.}{2016}]{Dijkstra2016}
{Dijkstra} M.,  {Gronke} M.,   {Venkatesan} A.,  2016, \mn@doi [\apj]
  {10.3847/0004-637X/828/2/71}, \href
  {https://ui.adsabs.harvard.edu/abs/2016ApJ...828...71D} {828, 71}

\bibitem[\protect\citeauthoryear{{Dopita} \& {Sutherland}}{{Dopita} \&
  {Sutherland}}{2003}]{Dopita2003}
{Dopita} M.~A.,  {Sutherland} R.~S.,  2003, {Astrophysics of the diffuse
  universe}

\bibitem[\protect\citeauthoryear{{Finkelstein} et~al.,}{{Finkelstein}
  et~al.}{2019}]{Finkelstein2019}
{Finkelstein} S.~L.,  et~al., 2019, \mn@doi [\apj] {10.3847/1538-4357/ab1ea8},
  \href {https://ui.adsabs.harvard.edu/abs/2019ApJ...879...36F} {879, 36}

\bibitem[\protect\citeauthoryear{{Harikane} et~al.,}{{Harikane}
  et~al.}{2018}]{Harikane2018}
{Harikane} Y.,  et~al., 2018, \mn@doi [\apj] {10.3847/1538-4357/aabd80}, \href
  {http://adsabs.harvard.edu/abs/2018ApJ...859...84H} {859, 84}

\bibitem[\protect\citeauthoryear{{Harikane} et~al.,}{{Harikane}
  et~al.}{2019}]{Harikane2019}
{Harikane} Y.,  et~al., 2019, \mn@doi [\apj] {10.3847/1538-4357/ab2cd5}, \href
  {https://ui.adsabs.harvard.edu/abs/2019ApJ...883..142H} {883, 142}

\bibitem[\protect\citeauthoryear{Hayes}{Hayes}{2019}]{Hayes2019}
Hayes M.,  2019, Lyman Alpha Emission and Absorption in Local Galaxies.
Springer Berlin Heidelberg, Berlin, Heidelberg, pp 319--398,
  \mn@doi{10.1007/978-3-662-59623-4_4}, \url
  {https://doi.org/10.1007/978-3-662-59623-4_4}

\bibitem[\protect\citeauthoryear{{Hayes}, {Schaerer}, {{\"O}stlin},
  {Mas-Hesse}, {Atek}  \& {Kunth}}{{Hayes} et~al.}{2011}]{Hayes2011}
{Hayes} M.,  {Schaerer} D.,  {{\"O}stlin} G.,  {Mas-Hesse} J.~M.,  {Atek} H.,
  {Kunth} D.,  2011, \mn@doi [\apj] {10.1088/0004-637X/730/1/8}, \href
  {http://adsabs.harvard.edu/abs/2011ApJ...730....8H} {730, 8}

\bibitem[\protect\citeauthoryear{{Higuchi} et~al.,}{{Higuchi}
  et~al.}{2019}]{Higuchi2019}
{Higuchi} R.,  et~al., 2019, \mn@doi [\apj] {10.3847/1538-4357/ab2192}, \href
  {https://ui.adsabs.harvard.edu/abs/2019ApJ...879...28H} {879, 28}

\bibitem[\protect\citeauthoryear{{Jiang} et~al.,}{{Jiang}
  et~al.}{2018}]{Jiang2018}
{Jiang} L.,  et~al., 2018, \mn@doi [Nature Astronomy]
  {10.1038/s41550-018-0587-9}, \href
  {https://ui.adsabs.harvard.edu/abs/2018NatAs...2..962J} {2, 962}

\bibitem[\protect\citeauthoryear{{Kennicutt}}{{Kennicutt}}{1998}]{Kennicut1998}
{Kennicutt} Robert~C. J.,  1998, \mn@doi [\apj] {10.1086/305588}, \href
  {https://ui.adsabs.harvard.edu/#abs/1998ApJ...498..541K} {498, 541}

\bibitem[\protect\citeauthoryear{{Leitherer} et~al.,}{{Leitherer}
  et~al.}{1999}]{Leitherer1999}
{Leitherer} C.,  et~al., 1999, \mn@doi [The Astrophysical Journal Supplement
  Series] {10.1086/313233}, \href
  {https://ui.adsabs.harvard.edu/#abs/1999ApJS..123....3L} {123, 3}

\bibitem[\protect\citeauthoryear{{Mas-Hesse} \& {Kunth}}{{Mas-Hesse} \&
  {Kunth}}{1991}]{MasHesse1991}
{Mas-Hesse} J.~M.,  {Kunth} D.,  1991, \aaps, \href
  {https://ui.adsabs.harvard.edu/abs/1991A&AS...88..399M} {88, 399}

\bibitem[\protect\citeauthoryear{Meyer, Laporte, Ellis, Verhamme  \&
  Garel}{Meyer et~al.}{2020}]{Meyer2020}
Meyer R.~A.,  Laporte N.,  Ellis R.~S.,  Verhamme A.,   Garel T.,  2020,
  \mn@doi [Monthly Notices of the Royal Astronomical Society]
  {10.1093/mnras/staa3216}, 500, 558

\bibitem[\protect\citeauthoryear{{Oke} \& {Gunn}}{{Oke} \&
  {Gunn}}{1983}]{Oke1983}
{Oke} J.~B.,  {Gunn} J.~E.,  1983, \mn@doi [\apj] {10.1086/160817}, \href
  {https://ui.adsabs.harvard.edu/abs/1983ApJ...266..713O} {266, 713}

\bibitem[\protect\citeauthoryear{{Osterbrock}}{{Osterbrock}}{1989}]{osterbrock}
{Osterbrock} D.~E.,  1989, \mn@doi [Annals of the New York Academy of Sciences]
  {10.1111/j.1749-6632.1989.tb50500.x}, \href
  {https://ui.adsabs.harvard.edu/abs/1989NYASA.571...99O} {571, 99}

\bibitem[\protect\citeauthoryear{{Oteo} et~al.,}{{Oteo}
  et~al.}{2018}]{Oteo2018}
{Oteo} I.,  et~al., 2018, \mn@doi [\apj] {10.3847/1538-4357/aaa1f1}, \href
  {http://adsabs.harvard.edu/abs/2018ApJ...856...72O} {856, 72}

\bibitem[\protect\citeauthoryear{{Ot{\'{\i}}-Floranes} \&
  {Mas-Hesse}}{{Ot{\'{\i}}-Floranes} \& {Mas-Hesse}}{2010}]{Oti2010}
{Ot{\'{\i}}-Floranes} H.,  {Mas-Hesse} J.~M.,  2010, \mn@doi [\aap]
  {10.1051/0004-6361/200913384}, \href {http://adsabs.harvard.edu/abs/2010A}
  {511, A61}

\bibitem[\protect\citeauthoryear{{Ouchi} et~al.,}{{Ouchi}
  et~al.}{2008}]{Ouchi2008}
{Ouchi} M.,  et~al., 2008, \mn@doi [\apjs] {10.1086/527673}, \href
  {http://adsabs.harvard.edu/abs/2008ApJS..176..301O} {176, 301}

\bibitem[\protect\citeauthoryear{{Ouchi} et~al.,}{{Ouchi}
  et~al.}{2010}]{Ouchi2010}
{Ouchi} M.,  et~al., 2010, \mn@doi [\apj] {10.1088/0004-637X/723/1/869}, \href
  {http://adsabs.harvard.edu/abs/2010ApJ...723..869O} {723, 869}

\bibitem[\protect\citeauthoryear{Robertson et~al.,}{Robertson
  et~al.}{2013}]{Robertson2013}
Robertson B.~E.,  et~al., 2013, \mn@doi [The Astrophysical Journal]
  {10.1088/0004-637x/768/1/71}, 768, 71

\bibitem[\protect\citeauthoryear{{Robertson}, {Ellis}, {Furlanetto}  \&
  {Dunlop}}{{Robertson} et~al.}{2015}]{Robertson15}
{Robertson} B.~E.,  {Ellis} R.~S.,  {Furlanetto} S.~R.,   {Dunlop} J.~S.,
  2015, \mn@doi [\apjl] {10.1088/2041-8205/802/2/L19}, \href
  {http://adsabs.harvard.edu/abs/2015ApJ...802L..19R} {802, L19}

\bibitem[\protect\citeauthoryear{{Rodr{\'\i}guez Espinosa}
  et~al.,}{{Rodr{\'\i}guez Espinosa} et~al.}{2020}]{RodriguezEspinosa2020}
{Rodr{\'\i}guez Espinosa} J.~M.,  et~al., 2020, \mn@doi [\mnras]
  {10.1093/mnrasl/slaa045}, \href
  {https://ui.adsabs.harvard.edu/abs/2020MNRAS.495L..17R} {495, L17}

\bibitem[\protect\citeauthoryear{{Salvador-Sol{\'e}}, {Manrique}, {Guzman},
  {Rodr{\'\i}guez Espinosa}, {Gallego}, {Herrero}, {Mas-Hesse}  \& {Mar{\'\i}n
  Franch}}{{Salvador-Sol{\'e}} et~al.}{2017}]{Salvador-Sole2017}
{Salvador-Sol{\'e}} E.,  {Manrique} A.,  {Guzman} R.,  {Rodr{\'\i}guez
  Espinosa} J.~M.,  {Gallego} J.,  {Herrero} A.,  {Mas-Hesse} J.~M.,
  {Mar{\'\i}n Franch} A.,  2017, \mn@doi [\apj] {10.3847/1538-4357/834/1/49},
  \href {https://ui.adsabs.harvard.edu/abs/2017ApJ...834...49S} {834, 49}

\bibitem[\protect\citeauthoryear{{Schaerer}}{{Schaerer}}{2003}]{schaerer2003}
{Schaerer} D.,  2003, \mn@doi [\aap] {10.1051/0004-6361:20021525}, \href
  {https://ui.adsabs.harvard.edu/abs/2003A&A...397..527S} {397, 527}

\bibitem[\protect\citeauthoryear{{Schenker}, {Stark}, {Ellis}, {Robertson},
  {Dunlop}, {McLure}, {Kneib}  \& {Richard}}{{Schenker}
  et~al.}{2012}]{Schenker2012}
{Schenker} M.~A.,  {Stark} D.~P.,  {Ellis} R.~S.,  {Robertson} B.~E.,  {Dunlop}
  J.~S.,  {McLure} R.~J.,  {Kneib} J.-P.,   {Richard} J.,  2012, \mn@doi [\apj]
  {10.1088/0004-637X/744/2/179}, \href
  {https://ui.adsabs.harvard.edu/abs/2012ApJ...744..179S} {744, 179}

\bibitem[\protect\citeauthoryear{{Sobral} \& {Matthee}}{{Sobral} \&
  {Matthee}}{2019}]{Sobral2019}
{Sobral} D.,  {Matthee} J.,  2019, \mn@doi [\aap]
  {10.1051/0004-6361/201833075}, \href
  {https://ui.adsabs.harvard.edu/abs/2019A&A...623A.157S} {623, A157}

\bibitem[\protect\citeauthoryear{{Stark}, {Ellis}, {Chiu}, {Ouchi}  \&
  {Bunker}}{{Stark} et~al.}{2010}]{Stark2010}
{Stark} D.~P.,  {Ellis} R.~S.,  {Chiu} K.,  {Ouchi} M.,   {Bunker} A.,  2010,
  \mn@doi [\mnras] {10.1111/j.1365-2966.2010.17227.x}, \href
  {http://adsabs.harvard.edu/abs/2010MNRAS.408.1628S} {408, 1628}

\bibitem[\protect\citeauthoryear{{Stark} et~al.,}{{Stark}
  et~al.}{2015}]{Stark2015}
{Stark} D.~P.,  et~al., 2015, \mn@doi [\mnras] {10.1093/mnras/stv1907}, \href
  {https://ui.adsabs.harvard.edu/abs/2015MNRAS.454.1393S} {454, 1393}

\bibitem[\protect\citeauthoryear{{Stark} et~al.,}{{Stark}
  et~al.}{2017}]{Stark2017}
{Stark} D.~P.,  et~al., 2017, \mn@doi [\mnras] {10.1093/mnras/stw2233}, \href
  {https://ui.adsabs.harvard.edu/abs/2017MNRAS.464..469S} {464, 469}

\bibitem[\protect\citeauthoryear{{Steidel}, {Adelberger}, {Shapley}, {Erb},
  {Reddy}  \& {Pettini}}{{Steidel} et~al.}{2005}]{Steidel2005}
{Steidel} C.~C.,  {Adelberger} K.~L.,  {Shapley} A.~E.,  {Erb} D.~K.,  {Reddy}
  N.~A.,   {Pettini} M.,  2005, \mn@doi [\apj] {10.1086/429989}, \href
  {https://ui.adsabs.harvard.edu/abs/2005ApJ...626...44S} {626, 44}

\bibitem[\protect\citeauthoryear{{Tilvi} et~al.,}{{Tilvi}
  et~al.}{2020}]{Tilvi2020}
{Tilvi} V.,  et~al., 2020, \mn@doi [\apjl] {10.3847/2041-8213/ab75ec}, \href
  {https://ui.adsabs.harvard.edu/abs/2020ApJ...891L..10T} {891, L10}

\bibitem[\protect\citeauthoryear{{Toshikawa} et~al.,}{{Toshikawa}
  et~al.}{2012}]{Toshikawa2012}
{Toshikawa} J.,  et~al., 2012, \mn@doi [\apj] {10.1088/0004-637X/750/2/137},
  \href {http://adsabs.harvard.edu/abs/2012ApJ...750..137T} {750, 137}

\bibitem[\protect\citeauthoryear{{Vanzella} et~al.,}{{Vanzella}
  et~al.}{2011}]{Vanzella2011}
{Vanzella} E.,  et~al., 2011, \mn@doi [\apjl] {10.1088/2041-8205/730/2/L35},
  \href {https://ui.adsabs.harvard.edu/abs/2011ApJ...730L..35V} {730, L35}

\bibitem[\protect\citeauthoryear{{Verhamme}, {Orlitov{\'a}}, {Schaerer},
  {Izotov}, {Worseck}, {Thuan}  \& {Guseva}}{{Verhamme}
  et~al.}{2017}]{Verhamme2017}
{Verhamme} A.,  {Orlitov{\'a}} I.,  {Schaerer} D.,  {Izotov} Y.,  {Worseck} G.,
   {Thuan} T.~X.,   {Guseva} N.,  2017, \mn@doi [\aap]
  {10.1051/0004-6361/201629264}, \href
  {http://adsabs.harvard.edu/abs/2017A%26A...597A..13V} {597, A13}

\bibitem[\protect\citeauthoryear{{Wilkins}, {Feng}, {Di-Matteo}, {Croft},
  {Stanway}, {Bouwens}  \& {Thomas}}{{Wilkins} et~al.}{2016}]{Wilkins2016}
{Wilkins} S.~M.,  {Feng} Y.,  {Di-Matteo} T.,  {Croft} R.,  {Stanway} E.~R.,
  {Bouwens} R.~J.,   {Thomas} P.,  2016, \mn@doi [\mnras]
  {10.1093/mnrasl/slw007}, \href
  {https://ui.adsabs.harvard.edu/abs/2016MNRAS.458L...6W} {458, L6}

\bibitem[\protect\citeauthoryear{{Wright}}{{Wright}}{2006}]{Wright2006}
{Wright} E.~L.,  2006, \mn@doi [\pasp] {10.1086/510102}, \href
  {https://ui.adsabs.harvard.edu/abs/2006PASP..118.1711W} {118, 1711}

\bibitem[\protect\citeauthoryear{{Yue} et~al.,}{{Yue} et~al.}{2018}]{Yue2018}
{Yue} B.,  et~al., 2018, \mn@doi [\apj] {10.3847/1538-4357/aae77f}, \href
  {https://ui.adsabs.harvard.edu/abs/2018ApJ...868..115Y} {868, 115}

\makeatother
\end{thebibliography}

\label{lastpage}
\end{document}